\documentclass[aps,twocolumn,10pt]{revtex4}

\usepackage{dcolumn}
\usepackage{graphicx}
\ifnum\lefthyphenmin<2\lefthyphenmin=2\fi
\ifnum\righthyphenmin<2\righthyphenmin=2\fi
\begin{document}

\title{Spontaneous Self-Assembly of Transcription Factor Based Gene Regulation Networks}

\author{ D. Balcan$^1$, A. Kabak\c c\i o\u glu$^2$, M. Mungan$^{3,4}$, and  A. Erzan$^{1,4}$\\}

\affiliation{$^1$Department of Physics, Faculty of Sciences and
Letters\\
Istanbul Technical University, Maslak 34469, Istanbul, Turkey}
\affiliation{$^2$Department of Physics, Faculty of Arts and Sciences, Koc University, 34450 Sariyer Istanbul, Turkey}
\affiliation{$^3$Department of Physics, Faculty of Arts and Sciences \\
Bogazi\c ci University, 34342 Bebek Istanbul, Turkey}
\affiliation{$^4$G\"ursey Institute, P.O.B. 6, \c Cengelk\"oy, 34680 Istanbul, Turkey}

\date{\today }

\begin{abstract}
We model the transcription factor based regulation network of
yeast using a content-based network model that mimicks the
recognition of binding motifs on the regulatory regions of the
genes. We are thereby able to faithfully reproduce many of the
topological features of the gene regulatory network of yeast once
the parameters of the yeast genome, in particular the distribution
of information coded by the ``binding sequences" within the
promoter regions is provided as input. The length distribution for
the promoter regions is fixed by comparing the k-core analysis of
the model network with that of yeast. Our results strongly point
to the possibility that the observed topological features are
generic to networks formed via sequence-matching between random
strings obeying certain length distributions.
\end{abstract}
\maketitle

\section{Introduction}
Development of new experimental techniques, such as DNA microarrays,
in the late 1990's~\cite{microarray,spellman} made a huge impact on
cell biology research. Such experiments generated a flood of
expression data for several well-studied single-cell species for which we
now have an almost complete list of not only the genes, but also the
interactions between them.
A cell is able to survive, grow and replicate due to the
collective actions of its genes. The adaptation and robustness of
its activities in a constantly changing environment is maintained
by the complex network of interactions between the genes.

The regulation of gene expression in a cell relies to a major
extent on dedicated proteins called transcription factors
(TFs).~\cite{Cell} These proteins come with a structure suited to
recognize and bind the DNA at specific locations called binding
sites.  The binding affinity of a TF on a certain DNA segment is
determined by the base sequence at the location. Each TF
preferentially binds certain regulatory sequences or binding
motifs, within the promoter regions (PRs) responsible for the
regulation of the gene. In the case of yeast, {\it Saccharomyces
cerevisiae}, a list of the binding motifs for more than 100 TFs
has  recently been provided.~\cite{Lee,Harbison} It was also
reported~\cite{Harbison} that the TF binding sites are located
with high probability within a window of several hundred bases
upstream of the transcription activation site (preceding the start
codon of the gene), although longer-distance action is also
possible.  In fact, the existence of a high-affinity binding motif
in a promoter region is a necessary but not sufficient condition
for TF-based expression regulation~\cite{Harbison}.  Moreover,
especially in eukaryotic cells, gene regulation relies on the
simultaneous action of multiple TFs.

We argue that the global features of the gene regulation network
depend very little on such details and are largely determined by
the distribution of the amount of shared information or content, 
that is required for the establishment of regulatory interactions.  
It may be conjectured that information sharing and its 
distribution is the basic organizing principle which is
responsible for the universality of the degree distribution of gene regulatory
networks across diverse species~\cite{Barkai}. 

In this paper we propose to model the transcription regulation network
of yeast using the ideas of the content-based model we
introduced earlier~\cite{Balcan,Mungan}.  We are able to
faithfully reproduce all the topological aspects of the gene
regulatory network of yeast when the parameters of the yeast genome,
in particular the distribution of information coded by the ``binding
sequences" of the regulatory segments, are given as input.  We compare
the ensemble of the resulting model networks with the data on the
yeast regulatory network available in different databases.

Gene regulatory networks can be naturally described as a directed
graph where the nodes are the genes. A directed edge from node A
to node B implies that the transcription factor produced by gene A
regulates the activity of gene B. Since the edges are directed,
one distinguishes the in-degree (the number of incoming edges),
the out-degree (number of outgoing edges) and the total degree of
a node, each with their own (possibly distinct) probability
distributions. These distributions serve as distinguishing
features of the network which a realistic model is expected to
reproduce. Further structural aspects of these networks are probed
by measures such as the clustering coefficient
$C(k)$~\cite{Dorogovtsev,Watts-Strogats98}, the degree-degree
correlation between connected
vertices~\cite{kk-correlation_colizza}, the ``rich-club
coefficient''~\cite{rich-club,rich-club_colizza}, or the $k$-core
decomposition~\cite{bollobas} recently
employed to predict new
interactions in various biological systems~\cite{protein_k-core,yeast_k-core,bader,amin,wuchty}.

This report is organized as follows: In Section \ref{modelSect} we
introduce our model, which we compare with the experimentally
determined yeast regulatory network in  \ref{SimSect}. A
discussion is provided in Section \ref{DiscSect}, while Section
\ref{MethodsSect} outlines our methods.

\section{The Model}
\label{modelSect}

The nodes of our model network correspond to genes. We
differentiate between genes which code for a Transcription Factor
(TF) and those which do not.  All genes are assumed to be possible
targets of regulation by one or more TFs.  Each node has a
sequence associated with it, representing the promoter region (PR)
through which the corresponding gene may be regulated. We pick a
given percentage of nodes (around 5\%, see Table I) at random, to
represent TF-producing genes. With each TF-producing node/gene we
also associate a second sequence, which stands for the binding
motif, which the TF recognizes and binds in the promoter region of
another gene.

We represent both the binding motifs and the PRs as random binary
sequences of variable length. The mechanism for establishing
connections between nodes of the gene regulatory network is given
by a string matching condition~\cite{Balcan,Mungan}, between the
binding motifs of the TF's and all possible uninterrupted
subsequences of the PRs. The (directed) network of regulatory gene
interactions is then obtained by connecting each TF-producing node
${\rm A}$ to all those nodes ${\rm B},\; {\rm B}^\prime,\;{\rm
B}^{\prime \prime} \ldots$ whose PRs contain the binding motif
associated with node ${\rm A}$. The amount of information coded in
these randomly generated binding motifs and promoter regions
constitutes the essential ingredient of our model and dictates the
overall topology of the resultant networks.

\begin{figure}[h]
\vspace*{0.0cm}
\includegraphics[width=7cm]{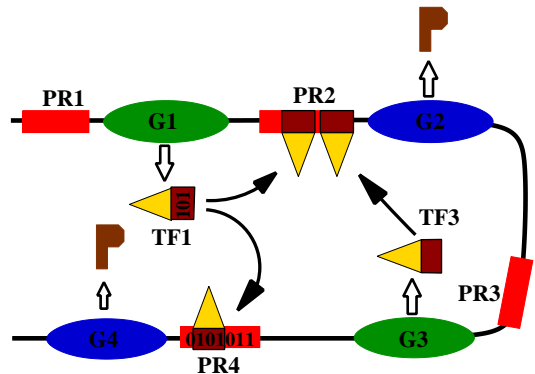}
\caption[]{The mechanism of interaction between the genes as envisaged
  in our model.  The genes are indicated by ellipses (green if
  TF-coding, blue otherwise), the transcription factors by triangles
  with the associated binding motif in the box underneath. Non-TF
  proteins are symbolized by the ``P'' shape, and the promoter regions
  (PR) upstream of each gene are shown as red boxes. Binding occurs if
  the binding motif matches a subsequence in the PR, as is the case
  here at PR4. PRs in the model are typically much longer than
  depicted here.}
\label{model}
\end{figure}

Experimentally determined TF binding motifs are typically short sequences
with a narrow length distribution, since a TF  selectively
binds 5-10 bases and not much more. A single TF can bind a range of
similar motifs, and the relative frequencies of the four bases at each
position within the motif contribute to the information exchanged in
the binding process.  The promoter regions (PRs) which lie in the
intergenic portions of the genome are typically longer and may
accommodate several binding motifs (as shown in Fig.~\ref{model}) to allow
graded and/or combinatorial regulation~\cite{Cell,Harbison}.

\begin{figure}[h]
\vspace*{0.8cm}
\includegraphics[width=6cm]{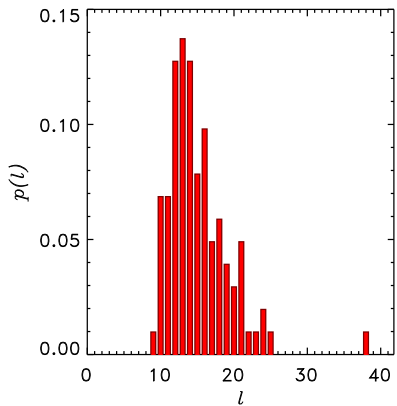}
%\rotatebox{270}{\scalebox{.4}{\includegraphics{Fig0.eps}}}
\caption[]{Distribution of the amount of bitwise information coded by each
  regulatory sequence recognized and bound by the 102 TFs in the yeast
  genome (compiled from the recently published data by Harbison et
  al.~\cite{Harbison}). This distribution is adopted as the length
  distribution of the random regulatory sequences (``binding motifs") in our model.}
\label{RS_dist}
\end{figure}

The bitwise length distribution of the model binding motifs was
derived from the yeast data provided by Harbison et al. in
\cite{Harbison}. The motifs were reported~\cite{Harbison} as letter
sequences comprising the symbols for the four bases \{ATGC\}, or
the symbols \{YMKRSW\} for incompletely specified bases, with the
corresponding lower case letters indicating a lower confidence level.
In order to account for such variations in the information content of
the motifs, we assigned two bits to each of the letters \{ACTG\}
appearing in the motif, signifying a high information content at
that position, and one bit otherwise. The
length of the bit sequence obtained in this way roughly corresponds
to the amount of shared information, measured by the Shannon
entropy~\cite{Shannon}, required for the binding of the TF.
Performing this calculation for each TF in~\cite{Harbison}, we obtain
the length distribution shown in Fig.~\ref{RS_dist}.

In choosing the length distribution of the promoter regions, about
which less is known, we are guided by the finding~\cite{Harbison}
that most of the probability for encountering a TF binding site is
contained within a window of 250 base pairs (bps) located
approximately 100 bps upstream of a gene. The PR length
distribution that we adopt within this range decays with a power
law  $p(l) \propto l^{-1-\mu}$, with $0\le\mu\le2$ after the
findings of Almirantis and Provata~\cite{Provata} for the lengths
of intergenic regions. We also assign a minimum length chosen to
coincide with the peak of the motif-length distribution shown in
Fig.~\ref{RS_dist}. Note that the 250 bps window does not double
as we move from the 4 letter alphabet to a binary one, because the
matching probabilities and the total number of positions at which
the TFs may bind are required to remain invariant under this
transformation.

The value of $\mu$ remains as the only adjustable
parameter in our model, and is determined by comparing the $k$-core
decomposition of the gene regulatory network of yeast as extracted
from experimental data (Table I) with our content-based network model,
as explained in the Methods section.

The collection of such model networks forms an ensemble whose
features are a direct consequence of the string-matching mechanism
and the length distributions. Clearly, each realization of the
model will result in a different collection of random PRs and
binding motifs, and hence a somewhat different network. These
features turn out to be strikingly distinct from those encountered
in random~\cite{erdos-renyi} or scale-free~\cite{Barabasi}
networks. We show below that the ``signatures'' of this ensemble
are shared by the yeast regulatory network.

\section{Results}
\label{SimSect}

Our purpose here is to show that the experimentally determined
features of the yeast regulation network follow closely those typical
of the ensemble defined by our model. The topological features we will
focus on are the following:

\begin{enumerate}
\item {\bf degree distribution} (in-, out-, and total): the
distribution of the number of connections of the nodes in a network.
\item {\bf clustering coefficient}: the modularity of the network.
\item {\bf degree-degree correlations}: average degree of the neighbors
of a node with degree $k$.
\item {\bf ``rich-club'' coefficient}:  a measure of the relative
connectivity among nodes whose degree is higher than a given number.
\item {\bf $k$-core structure}: the hierarchical structuring in the network
\end{enumerate}

The precise definition of these quantities is given in the Methods
section.

Here we will report the comparison of our results with the most
recent Yeastract~\cite{Nucleicacids} data.  Analogous comparisons
with each of the data sources listed in Table~\ref{tabyeast} yield
similar results (see Supplementary Material) showing that our
conclusions are consistent with all the different data sets
available.

In order to compare our results with the available data we
generate an ensemble of realizations, with an average of $N_G =
6000$ genes in total, 4167 of which contribute to the network on
the average. Out of these, 202 (making up  \% 4.8 of the genes)
are TF-coding genes, taking part in a total of 14365 interactions,
again on the average.  The corresponding values for the yeast
regulatory networks reported in the publicly available data bases
are given in Table~\ref{tabyeast}.

The total degree distribution is obtained by ignoring the
directionality of the interactions and is different from the
superposition of in- and out-degree distributions. In
Fig.~\ref{degree-dists}a,  Yeastract data for the degree
distribution  is shown on top of a scatter plot obtained by
superposing the results from 100 artificial model genomes
independently generated according to the rules described in
Section \ref{modelSect}. In Fig.~\ref{degree-dists}b, we exhibit
the in-degree distribution obtained from the Yeastract data, and
the corresponding scatter plot.

\vspace{0.0cm}
\begin{table}
\caption[]{The number of interacting genes, TFs, and interacting pairs that appear
in the yeast regulatory network as obtained from different sources.}
\begin{tabular}{l|c|c|c}
\hline Source & Genes & TFs & Interacting Pairs \\ \hline \hline
Fraenkel Lab\footnote{http://fraenkel.mit.edu/Harbison/release\_v24/bound\_by\_factor/} & 2884 & 102 & 6441  \\ \hline
Yeastract\footnote{http://www.yeastract.com}
 & 4252 & 146 & 12530 \\ \hline
Luscombe et al.\footnote{http://sandy.topnet.gersteinlab.org/index2.html}
 & 3459 & 142 & 7071 \\ \hline
K\i rdar et al.\footnote{private communication} & 3763 & 180 & 9135 \\
\end{tabular}
\label{tabyeast}
\end{table}

The out-degree distribution of the yeast and model networks exhibits a rather
large scatter of points due to the relatively small number of TFs.
Comparing with the scatter plot obtained from 100 realizations, we find again
that the actual yeast data falls within the boundaries set by the model ensemble
(Fig.~\ref{degree-dists}c).

In Fig.~\ref{coefficients}, we report the three topological
coefficients, the clustering coefficient, the degree-degree
correlation and the ``rich-club'' coefficient, that go beyond
degree-distributions in characterizing the network. The agreement is extremely good;
in particular, the shoulder observed in the ``rich-club''
coefficient in Fig.~\ref{coefficients}(c), a feature common to both
gene-regulation and protein-protein interaction networks
\cite{kk-correlation_colizza}, is captured accurately in our model.

The agreement observed with the Yeastract data is not
source-specific, as can be seen from a comparison of the
topological properties of our model networks, with those
%for the yeast networks as
obtained from the different sources listed in Table
\ref{tabyeast}. (see Supplement)

Finally, in Fig.~\ref{k-core}, left, the $k$-core analysis of the
model network is shown, which should be compared with that of the
Yeastract data on the right. The $k$-core analysis provides a much
more stringent characterization of a network than the other single
topological features considered above. To give an idea of the
sensitivity of the $k$-core analysis to the structure of the
network,  let us point out that, under a shuffling of the edges of
the network keeping the degree of each node fixed,  the typical
value of the maximum number of $k$-cores, $k_{\rm max}$, becomes
29 rather than 9 as observed in both the real yeast regulatory
network and the model (see Supplement).

\begin{widetext}

\begin{figure}
\vspace*{0.0cm}
\includegraphics[width=17.0cm]{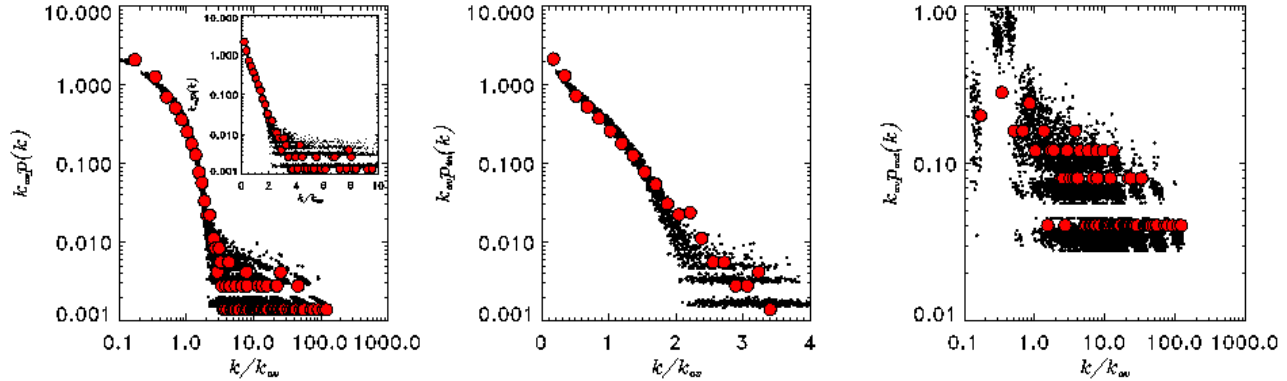}
\caption[]{Degree distributions extracted from the
  Yeastract~\cite{Nucleicacids} data (red circles), superposed on the
  corresponding degree distributions of 100 realizations of the model
  network (black dots). From left to right, a) The total degree distribution
  with an inset showing a log-linear plot for $k/k_{\rm av} \le 10$,
  where one may observe that both the model and the data points almost
  fall on a straight line. b) The in-degree distribution
  plotted on a semi-logarithmic scale. c) The out-degree distribution
  plotted on a log-log scale. The axes are  scaled by the average
  total degree in order to factor out sample-to-sample fluctuations in the network
  size.}
\label{degree-dists}
\end{figure}

\begin{figure}
\vspace*{0.0cm}
\includegraphics[width=17cm]{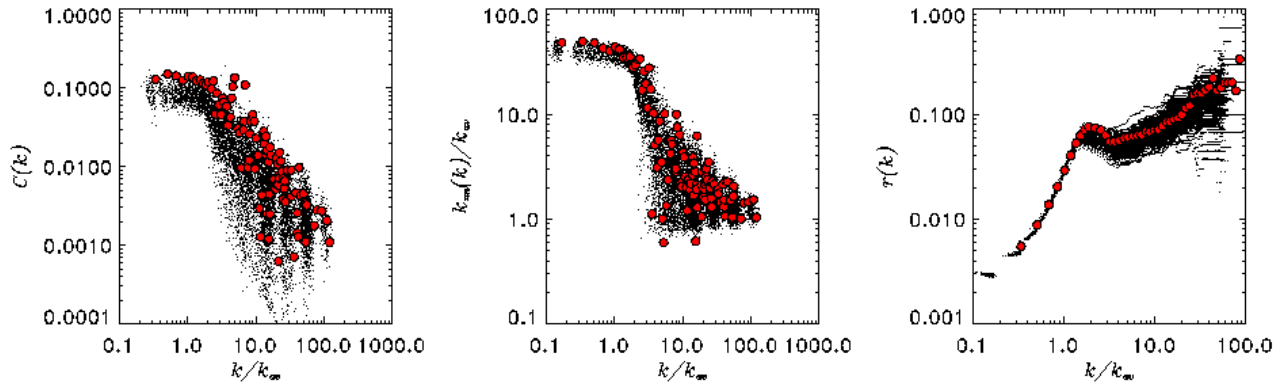}
\caption[]{ Comparison of a) the clustering coefficient $c(k)$, b)
the degree-degree correlations between neighboring nodes
$k_{nn}(k)$, and c) the rich-club coefficient $r(k)$, from left to right, for $100$
realizations of the model (black dots) and the Yeastract data (red
circles).} \label{coefficients}
\end{figure}

\end{widetext}

\section{Discussion}
\label{DiscSect}

The close structural similarity between the model and the real yeast regulatory network, with respect to a diverse set of criteria, shows that they are
part of the same statistical ensemble of networks, formed by random strings connected by the sequence matching rule.

The sequence matching rule could more generally be viewed as an
information-theoretical constraint, where the interaction between two genes
requires the fulfillment of a set of conditions which we
symbolically represent as the matching of two random sequences. The more
stringent the prerequisites of the interaction, the longer is the random
``binding motif" that is to be matched.
The length of the PR establishes the size of the phase space in which the motif is to be sought.
The properties of the network are then determined by the distributions obeyed by the lengths of the binding motifs as well as the promoting regions.

Interpreted within this information-theoretical framework, our model has sufficient
generality to accommodate other interactions based on lock-and-key mechanisms, such as protein networks, where the
interactions are dictated by certain steric and chemical conditions.

The topological features of the networks investigated here and
shown to be shared by the yeast regulatory network
strongly point to the possibility that these networks did not have to
be assembled from scratch, but rather emerged spontaneously,
given any sufficiently long linear code.
This proposition by no means minimizes the role of evolutionary pressures on such networks; instead, it
suggests that a network with essentially the current topology could have provided
a starting point for further fine-tuning. As a case in point, it has recently been demonstrated that evolution under duplication and divergence~\cite{Wagner} may leave the topological features of such networks essentially invariant~\cite{sengun}. Such a perspective will hopefully bring us a step
closer to envisioning how complex structures may have
come into existence, by shifting some of the load from the shoulders of
evolution onto the laws of probability.

\begin{widetext}

\begin{figure}
\vspace*{0.0cm}
\includegraphics[width=17cm]{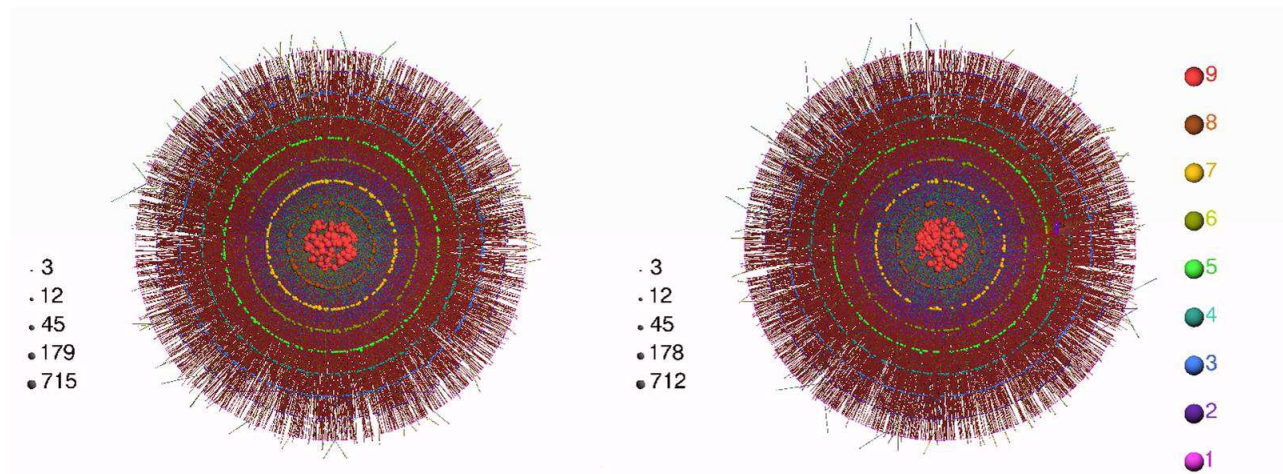}
\caption[]{Left: The $k$-core decomposition of a single realization of our model
  network obtained with the visualization tool lanet-vi~\cite{lanet-vi}.
  The length distribution exponent of the PR sequences has been
  adjusted to $\mu=0.1$ to optimize the similarity with the $k$-core
  distribution of the Yeastract data (Right). Dots represent the nodes
  of the network, while edges between nodes depict connections. Nodes
  belonging to different $k$-shells are indicated by different colors
  (on the right hand side) and are arranged around concentric circles,
  whose average radius decreases with k. In particular, a node of a
  given shell is placed just inside (outside) the corresponding circle, if it
  is preferentially connected to lower (higher) k-shells. The size of
dots indicate the degree of the respective nodes; see legends to the left of the figures.
}
\label{k-core}
\end{figure}

\end{widetext}

\section{Methods}
\label{MethodsSect}

The degree $k$ of a node is the number of edges connected to it.
When the graph is directed, one distinguishes in-, out-, and
total-degrees of a node, with their corresponding distributions.
In the measures below we have ignored the directionality of the
network.

The clustering coefficient is given by the formula:
\[ C_i = \frac{\Delta_i}{k_i(k_i-1)/2}\;,\]
where $\Delta_i $ is the number of triangles that contain node $i$.
The quantity $C(k)$ plotted
in Fig.~\ref{coefficients} is the average of $C_i$ over the nodes with
degree $k$.

The degree-degree correlation function $k_{nn}(k)$  is
\[
k_{nn}(k) = \sum_{k^\prime} k^\prime p(k^\prime \vert k),
\]
where $p(k^\prime \vert k)$ is the conditional probability that a node with
degree $k$ is connected to a node with degree $k^\prime$.

The``rich-club'' coefficient \cite{rich-club,rich-club_colizza}
$r(k)$ is the total number $e_{>k}$
of edges connecting nodes with degree greater than $k$, normalized by the
maximum possible number of such connections,
\[
r(k) = \frac{2e_{>k}}{N_{>k} (N_{>k} -1)},
\]
where $N_{>k}$ is the total number of nodes with degree greater than
$k$.

The $k$-core decomposition performs a successive pruning on the least
connected vertices of a network~\cite{bollobas}. At each step one
removes all nodes with a degree less than $k$ along with their edges and
continues in this manner until all
nodes have at least degree $k$. The remaining nodes constitute
the $k$ core. Next, $k$ is incremented by one, and the process
is repeated until no nodes are left. The $k$-shell is defined as
the set of nodes that belong to the $k$-core, but not the $(k+1)$-core.

Once the shape of the TF length distribution, the width of the PR region, as well
as the functional form of its distribution have been fixed through the available
biological data, the only remaining adjustable parameter in our model is the exponent
$\mu$ of the power law distribution of PR lengths, $p(l) \propto l^{-1-\mu}$. The $k$-core decomposition turns
out to provide the most detailed and stringent topological characterization of the
network, with both the total number of shells, and the distribution of the nodes
over the shells, being contained in the $k$-core plots (see Fig.\ref{k-core}). The
$k$-core plots also incorporate such qualitative features  as inter- and
intra-shell connectivity. We have therefore used qualitative and quantitative
comparison of the $k$-core plots for the Yeastract and the model network to determine $\mu$.
The best agreement was obtained for $\mu=0.1$.  Once $\mu$ has been fixed, no further
adjustment is needed in order to obtain the extremely close matching that is found
between the degree distributions, clustering coefficients, degree
correlations and the rich-club coefficient, as displayed in
Figs.~\ref{degree-dists} and ~\ref{coefficients}.

We cannot rule out the possibility of obtaining similar agreement between our model and the real genomic network with respect to the features considered here, for a different choice of the functional form of the length distribution for the PR sequences, once more determining an adjustable parameter from a  comparison of the $k$-core plots. However, the present choice seems to be the only reasonable one within the physical constraints and the available information.

\section{Acknowledgments}

We would like to thank Bet\"ul K\i rdar and Beste K\i n\i ko\u glu
for the use of their data and useful discussions. It is a pleasure
to thank Alessandro Vespignani and Ignacio Alvarez-Hamelin for
bringing $k$-core analysis to our attention, and for the use of
their web-based $k$-core analysis tool. AE would like to thank
Tam\'as Vicsek and Andr\'as Czir\'ok for a useful discussion and
is grateful for partial support from the Turkish Academy of
Sciences.

\begin{widetext}
\newpage

\newpage
%\bigskip
{\bf Supplementary Material 1}

{\bf Comparison with yeast data from different data bases}

%\begin{widetext}

\begin{figure}[h]
\vspace*{0.0cm}
\includegraphics[width=12cm]{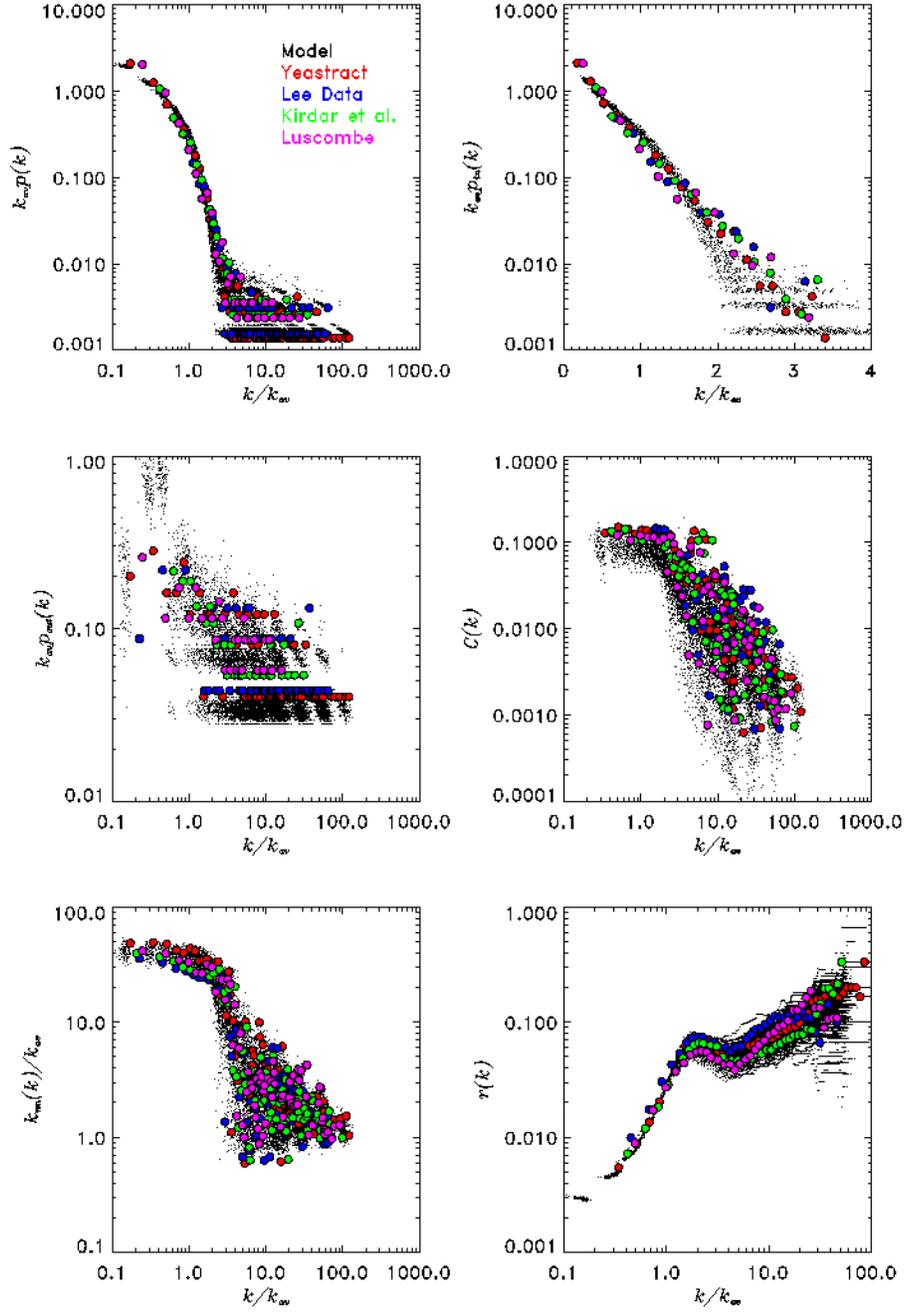}
\caption[]{The network statistics extracted from the sources
listed in Table~\ref{tabyeast} superposed on the simulation results corresponding to
100 realizations of the model network (black dots). The agreement is extremely good with all of these sets of data,
which almost completely cover, but do not exceed the phase space
of our model. (Black, red, blue, green yellow and maroon correspond to the model, Yeastract,
Fraenkel Lab, K\i rdar and Luscombe data respectively).
}
\label{supp_fig1}
\end{figure}

\newpage

{\bf Supplementary Material 2}

{\bf Comparison with Randomized Networks}

To double check the significance of our other results, we also
compared the clustering coefficients, the degree-degree correlations
and the rich-club coefficients of the Yeastract data
with those obtained after the randomly reconnecting the edges of the network while keeping the degree of each node fixed.  In this process, the directionality of the bonds is ignored.
The comparison of the topological coefficients of the randomized yeast and randomized model networks with that of the yeast network, as shown in Fig.~(\ref{randomized}), confirm that
the observed agreement between the yeast and models networks is not spurious.

%\begin{widetext}

\begin{figure}[h]
\vspace*{0.0cm}
\includegraphics[width=17cm]{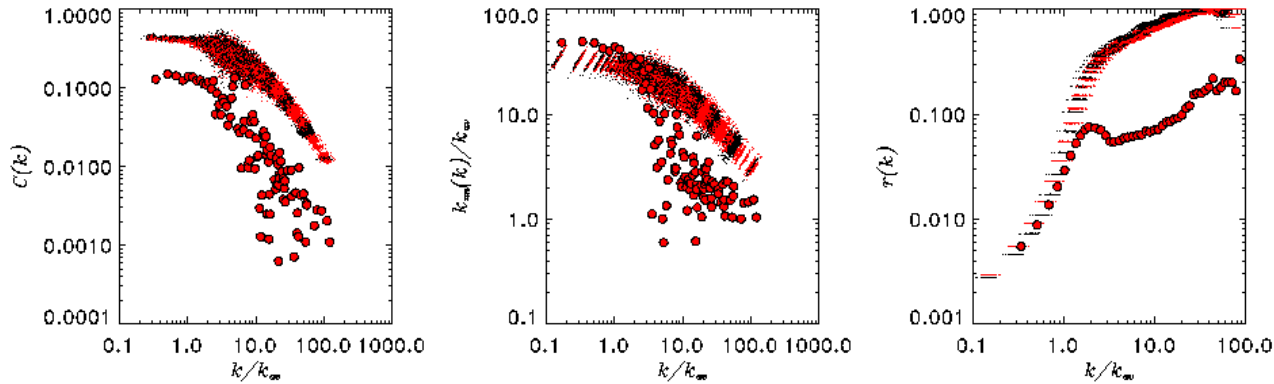}
\caption[]{a) The clustering coefficient, b) the degree-degree
correlations between neighboring nodes, and c) the rich-club
coefficient of Yeastract data (red circles) compared with the results
for the same obtained by randomizing the Yeastract data (red dots) and
randomizing a realization of the model network (black dots),
keeping the degrees of the individual nodes, and thereby the degree distributions, fixed.}
\label{randomized}
\end{figure}
%\end{widetext}

%\newpage

In Fig.~\ref{k-core-random} we display the effect of performing the same randomization procedure as described above, on the $k$-core plots.  It is instructive to note that while in the yeast and model networks,  a large fraction of connections is between nearby shells, the situation is reversed in the randomized networks, where there is a high degree of intra-shell connectivity as can be seen from Fig.~\ref{k-core}.

\begin{figure}[h]
\vspace*{0.0cm}
\includegraphics[width=17cm]{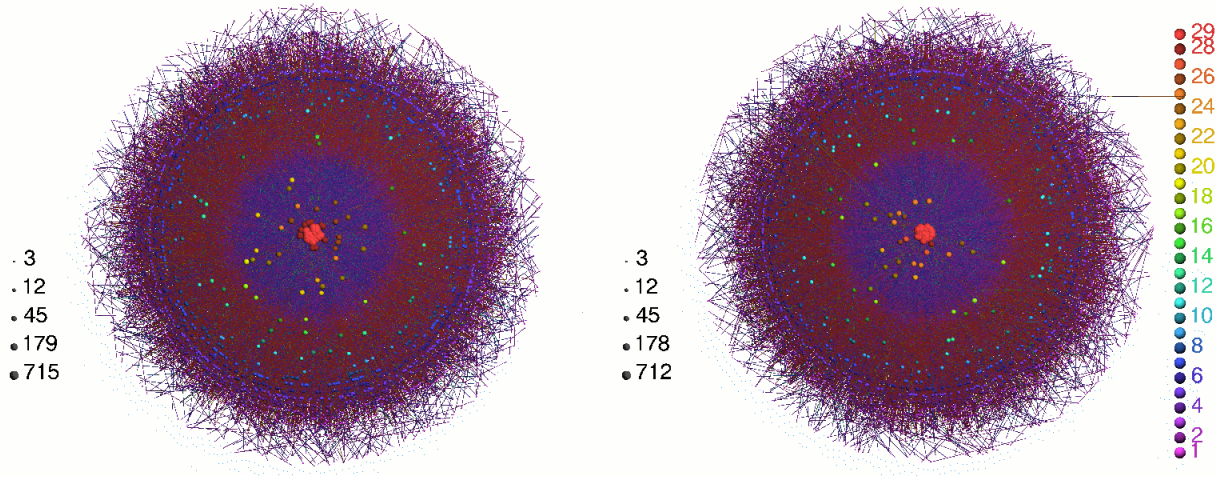}
\caption[]{The  $k$-core analysis of the randomized versions of the  model (left panel) and Yeastract (right panel) networks yield results that differ quantitatively and qualitatively from the originals.  The number of shells have gone up to 29 from 9, and the much higher intra-shell rather than inter-shell connectivity (as can be seen by following the edges) indicates that the hierarchical nature of the yeast network, which  is faithfully  reproduced by the model, is destroyed by the randomization process.
}
\label{k-core-random}
\end{figure}

\newpage

{\bf Supplementary Material 3}

{\bf The k-core structure of the Balcan-Erzan and Barabasi-Albert
Networks}

In Fig.~\ref{k-core-others} we show the k-core structure of the Balcan-Erzan~\cite{Balcan}
and Barabasi-Albert~\cite{BA} network, as models for complex networks. Note the absence
of well-defined hierarchical structures.

\begin{figure}[h]
\vspace*{0.0cm}
\includegraphics[width=17cm]{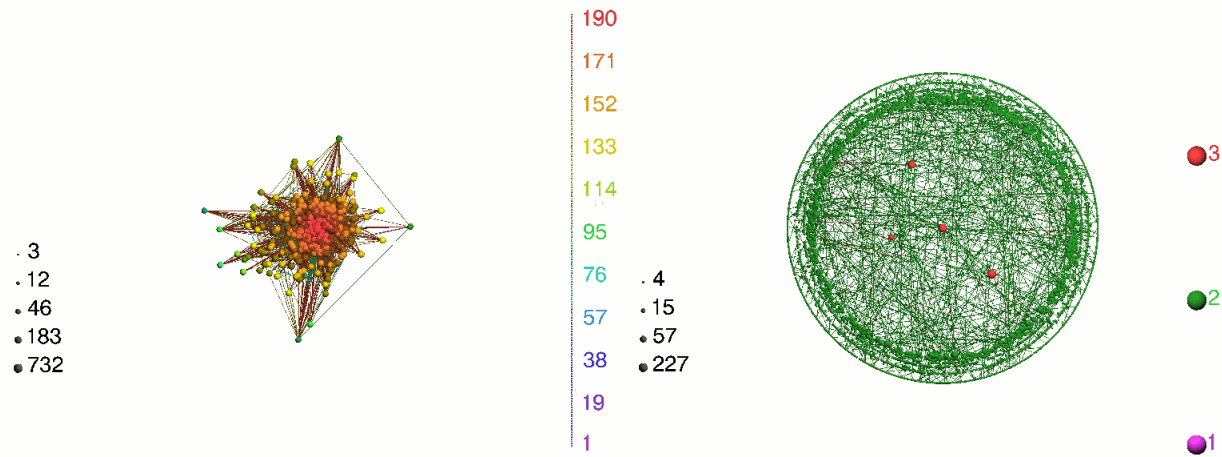}
\caption[]{The $k$-core analysis of the content-based network of
Balcan and Erzan~\cite{Balcan} (left panel) and the
Barabasi-Albert (BA) model~\cite{BA}. In the left panel, the total
length of the single sequences associated with all of the nodes is
$L = 15000$. The individual sequences obey the length distribution
$p(l) \propto q^l$, with $q = 0.95$. The BA model network (right
panel) has  5000 nodes, and is built by starting from a fully
connected four-cluster and adding nodes with two edges at a time.
In the $k$-core plot for the latter, only \% 5 of the edges are
shown for better visibility. } \label{k-core-others}
\end{figure}

\newpage
%\bigskip

{\bf Supplementary Material 4}

{\bf Ranking of overlapping sets of regulated genes and motif inclusion}

We here report  a statistical fact in support of
the basic assumption underlying our model. The matching condition we
employ dictates a certain correlation between the sets of regulated
genes by each TF: if the binding motif of a TF (A) is embedded in that
of a TF (B), then the set of genes \{G$_i$\}$_{\mbox{B}}$ regulated by
TF$_{\mbox{B}}$ in our model is a subset of \{G$_i$\}$_{\mbox{A}}$. A
similar investigation of the yeast databases listed below reveals that
the top 50\% of the TF pairs related by the motif inclusion relation
above, rank in the top 3\% when all the TF pairs are listed according
to the overlap of their \{G$_i$\} sets. The actual ranking
of the TF pairs obtained among all possible pairs of 102 TFs with
known binding motifs is shown in Fig.~\ref{TFcorr}.

\begin{figure}[h]
\vspace*{0.8cm}
\includegraphics[width=8cm]{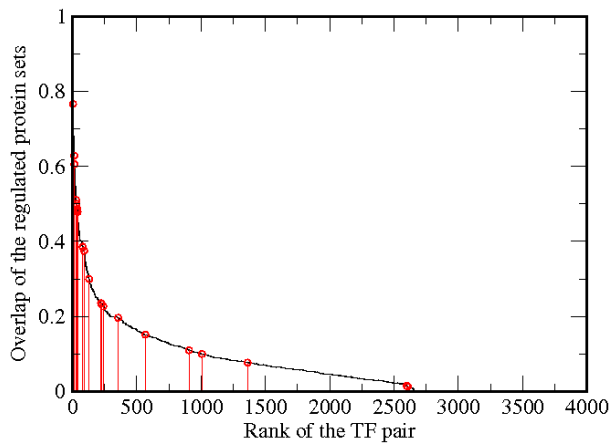}
\caption[]{Correlation between the sets of proteins regulated by
  the TFs with similar binding motifs.  The vertical axis is the
  percentage overlap of the two sets of genes regulated by an
  arbitrary pair of TFs, which are ranked on the horizontal axis
  according to their overlap.  The red vertical lines mark those pairs
  of TFs that are also related by binding motif inclusion. The
  accumulation of the red lines to the left of the graph is indicative
  of the correlation described in the text.}
\label{TFcorr}
\end{figure}

On the other hand, the more straighforward expectation that TFs with
short binding motifs should regulate more genes is not verified by the same
data. This curious fact probably points to certain sequence
correlations arising from the  duplication and divergence processes
~\cite{Wagner}
that distort the occurance statistics of the binding motifs in PRs. Note that
the result in Fig.~\ref{TFcorr} is robust to such deviations from the
unbiased probabilities for the occurance of different strings.

\end{widetext}

\end{document}